# A Feasibility Study on Usability and Trust among Population Groups of a Medical Avatar Supported by Large Language Models with Retrieval Augmented Generation

Roel Boumans[1], Lisa Cramer[1], Sascha van de Poll[2], Henria Vermeulen[2]

[1] Behavioural Science Institute, Radboud University, Nijmegen, Netherlands
[2] Canisius Wilhelmina Hospital, Nijmegen, Netherlands
`roel.boumans@ru.nl`

**Abstract.** Healthcare professionals have limited time to support patients and their relatives, but their information needs are high. Therefore, the Radboud University together with the Canisius Wilhelmina Hospital hospital developed a speaking virtual human avatar which, contrary to many avatars, uses a Large Language Model (LLM) enhanced with Retrieval Augmented Generation (RAG). The RAG technique enables medical information supplied by the hospital to be utilized during interactions, rather than generic LLM information. Two videos were produced, one presenting a patient-avatar interaction regarding a total hip surgery, and another one presenting an interaction between a relative of a patient and the avatar concerning postoperative delirium. A survey was conducted among adults over 40 from the Netherlands, the UK and the USA to study the effects of gender, country and education level on usability and trust, which are important factors for avatar acceptance. Participants watched videos, imagining themselves as the patient (video 1) or relative (video 2), and rated the constructs on a 7-point Likert scale (0-6). 165 persons (MeanAge=51.6, SDAge=8.9, Male=80, Female=85) completed the survey. In the patient role, participants scored the usability as $M$=4.61 ($SD$=0.97) and trust as $M$=3.92 ($SD$=1.10), all above the mean scale value. In the role as relative to the patient, participants scored usability as $M$=4.64 ($SD$=1.08) and trust as $M$=4.31 ($SD$=1.06). No effects were found of gender, country and education level.

**Keywords:** Virtual agent, avatar, dialogue manager, medicine.

## 1 Introduction

Many high-income countries are faced with an aging population that provides significant health challenges [1]. By 2050, 57 countries will have more than 30% of its population aged 60 and older [1]. At the same time the number of available healthcare professionals is decreasing [2]. The Organization for Economic Co-operation and Development (OECD) projects a shortage of 400,000 doctors and 2.5 million nurses by 2030 in OECD countries [3]. This combination of factors results in a high workload for the remaining healthcare professionals. Faced with heavy workloads, healthcare professionals must make choices between caring activities they find essential and caring activities that could be possibly omitted, often referred to as "care left undone" [4]. The most commonly reported types of "care left undone" involve comforting or talking to patients, educating patients and developing or updating nursing care plans [5]. The prevalence of "care left undone" is high (55-98%) [4]. At the same time, patients do like to talk to their formal caregivers in search of information about their situation [6].

Kynoch et al. (2019) conducted a scoping review of the information needs and information seeking behaviors of patients and families from healthcare providers and found that information retrieval preferences varied much across context [7]. Patients increasingly seek information on the Internet, e.g. on websites and apps, and use it in their contacts with healthcare professionals [8, 9]. Many websites are equipped with chatbots, and these can also be used to search for medical information [10]. Chatbots can provide patients with 24/7 access to health information, such as symptom assessment and supportive information [10]. However, a chatbot agent requires a typing interface, and this is not convenient in all situations and for all patients. Voice-based interactions with conversational agents offer several advantages over text-based interactions [11]. Voice interfaces are perceived as more efficient, enjoyable, and satisfying, while requiring less cognitive effort [11]. They facilitate more fluid conversations, increased turn-taking, and faster-paced discussions [12].

In response to the constraints on healthcare professionals' time to assist patients and their families, researchers at the Radboud University, in collaboration with Canisius Wilhelmina Hospital, have conceived and developed a graphical speech-based avatar. In this study, avatars refer to virtual humanoid characters that enable users to engage in verbal communication on specific topics. In other studies, they are also referred to as intelligent virtual agents, conversational agents or chatbots. The avatar is displayed on a 24" touchscreen and mounted on top of a computer-on-wheels (COW) platform for easy transportation within a hospital (**Fig. 1**). This avatar uses speech-to-text and text-to-speech functionalities, and its behavior is managed by a dialogue manager programmed in Python. A Retrieval Augmented Generation (RAG) enhanced LLM is used to answer medical questions from patients and relatives.

This solution aims to provide round-the-clock support to patients with hospital-provided medical information during their hospital stay. Additionally, it is easily transportable within the hospital and can be conveniently positioned next to patients' beds.

The concept of using an avatar (often also referred to as intelligent virtual agent or conversational agent or voicebot [13, 14]) for medical purposes in hospitals is not entirely new. Studies from related work focus clinical interviewing, medical training with virtual patients, and coaching patients for rehabilitation or life style improvement.

Concerning clinical interviewing, DeVault et al. (2014) designed SimSensei Kiosk to interview people with post-traumatic stress disorder (PTSD) [15, 16]. In more recent work, a task-oriented empathic chatbot, EmoBot, using RAGs was developed by Shayaninasab et al. (2024) for a mental healthcare patient intake scenario [17]. The acceptability of an embodied conversational agent for insomnia screening was evaluated by Menard et al. (2024) [18].

Medical training was studied by Rouleau et al. (2022), who used virtual patient simulation to improve nurses' relational skills, and by Galland et al. (2024) in the context of motivational interviewing [19, 20].

Regarding patient coaching, Retevoi et al. (2024) evaluated a conversational agent for stroke rehabilitation [21], Steenstra et al. (2024) used a virtual agent for alcohol use counseling [22], and Fallah et al. (2024) studied agenda setting in virtual health counseling sessions [23]. Valtolina et al. (2024) designed a conversational agent to empower older adults in active aging [24]. Archana et al. (2025) evaluate several chatbots and propose DocBot, a LLM powered medical chatbot which is specifically optimized for biomedical applications through Retrieval Augmented Generation (RAG), and evaluated its answers but without real user interaction and the use of voice [25].

However, studies explicitely evaluating the use of a LLM and RAG supported avatar for medical applications were found to be scarce. This study therefore builds upon this aforementioned work amongst others by combining the concept of a graphical virtual healthcare professional as designed by DeVault et al. [15], and use of RAG supported chatbots as designed by Shayaninasab et al. and Archana et al. [17, 25].

Our research question therefore was "What are the usability of and the trust in an avatar in a medical context when this avatar is supported by a Retrieval Augmented Generation enhanced Large Language Model?".

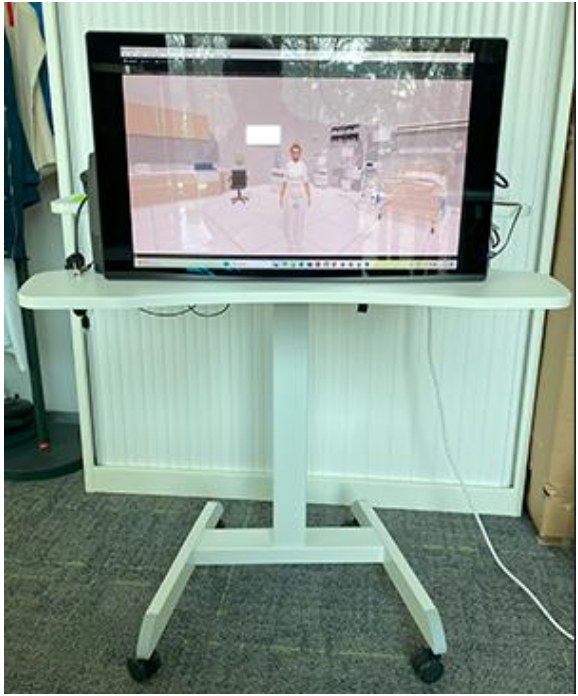

**Fig. 1.** Avatar on a Computer-On-Wheels platform

## 2    Theory and hypotheses

### 2.1    The Social Identity Theory

The Social Identity Theory tries to explain intergroup behavior [39]. Individuals tend to identify with certain social groups. They categorize themselves and strive to maintain a positive self-image. For this reason, they compare the social group with which they identify, the 'ingroup', with the social group with which they do not identify, the 'outgroup'. The evaluation of one's own group is determined in relation to specific other groups by social comparisons in terms of characteristics. Positively deviating comparisons between in-group and out-group result in high esteem and higher self-esteem, while negatively deviating comparisons between in-group and out-group result in low esteem and reinforced prejudices towards the out-group. Within this study, the participant group is analyzed per country of residence, gender, and education level, to study if any differences between these social groups are found that warrant further study.

Folk et al. (2025) found evidence for cultural differences in attitudes toward socially bonding with conversational agents [40]. In two studies with East Asian, European and American adults, they found differences in comfort, approval and connectedness between the user and the agent [40]. Scholz et al. (2023) performed a systematic review of social identity formalizations across social agents in different domains [41]. They found many models that tried to explain the social influence on user groups. Country of residence, being an indicator of the cultural background, could be of influence on the opinions of the users towards the avatar.

Regarding gender of the user, Eyssel et al. (2012) showed that the participant's gender affects the anthropomorphism and evaluation of robots [42]. They found, amongst others, that a short-haired male robot was perceived as more agentic than a long-haired female robot, while this female robot was perceived as more communal than her male counterpart [42]. Powers et al. (2005) studied how male or female users discussed with robots about romantic dating norms, and how this discussion affected their opinion on the robot [43]. Their results show that most ratings were the same across robot gender and subject gender, and in moderate ranges of the scale [43]. However, men's ratings of the female robot were significantly lower than women's ratings of either gendered robot and lower than men's ratings of the male robot [43].

The education level of participants is generally examined as a control or demographic variable and is not usually associated with a specific construct from Social Identity Theory. This raises the question if education level has an effect on the perceived usability and trust of the avatar.

The cited studies report differing impacts for these three factors. Thus, this study takes a conservative stance on the hypotheses regarding country of residence, gender, and education level. The first hypothesis is as follows:

**Hypothesis 1**: There is no difference between the participant groups and roles in terms of country of residence, gender, and education level in both usability and trust in the avatar.

## 2.2     The Source Credibility Theory

The Source Credibility Theory posits that the effectiveness of communication depends largely on how credible the audience perceives its source to be. In the context of this study, the avatar acts as the source whose credibility is evaluated. According to the theory, key factors such as competence, character, and sociability shape the perception of credibility [26, 27].

The Source Credibility Theory has been used by several scholars as background for studying communication with conversational agents. For example, Wienrich et al. (2021) used the Source Credibility Theory as background for evaluating the trust in voice assistants in the context of healthcare [28]. They recruited forty students who were randomized over two groups and were asked to answer questions from a voicebot labelled as "specialist" or "generalist". Their study results revealed that perceiving the voicebot as a specialist resulted in higher trustworthiness of the VA [28]. Liu et al. (2022) investigated how perceived source expertise influence usage intention of health chatbots [29]. They found in an experiment with 260 participants (91% with age < 35) that source expertise had no effect on usage intention [29]. Schillaci et al. (2024) found in a study with 1147 users that the relationship between healthcare chatbot personality and usability is mediated by the chatbot's perceived credibility [30]. They also found that, when there is a mismatch between chatbot's gender and its stereotypical descriptive property (i.e., competence for female chatbots; warmth for male chatbots), the chatbot is more credible, satisfying, and inviting to use [30].

Based on these findings, this study measures perceived credibility by evaluating usability and trust. If users find the avatar competent, relatable, and authoritative—qualities enhanced through visual cues like hospital logos, uniforms, and a hospital setting—they are more likely to find the avatar easy to use (high usability) and to believe the information it provides (high trust). The avatar's tone of voice, the clarity of its interface, and the fact that its responses are informed by real hospital brochures further reinforce these qualities.

Thus, the Source Credibility Theory directly underpins the study's focus on usability and trust: both variables are practical expressions of the avatar's credibility in the eyes of patients and caregivers. When participants perceive the avatar as a credible healthcare professional, their reported usability and trust in the avatar are expected to be higher. This theoretical framework justifies evaluating these particular variables as indicators of the avatar's effectiveness in delivering medical information.

Avatars can not only be used by the patients themselves, but also by their relatives. Likewise, these relatives are also concerned about the illness of their loved ones. When judging received healthcare, patients focus on relief of suffering and interpersonal kindness and may downplay negatives because they are in a dependent situation. Relatives, however, may act more as advocates, noticing systemic shortcomings, poor communication, or unmet needs. Some studies show that relatives are more critical than patients [31–33]. At the same time, some paired studies show the reverse or mixed results [34, 35]. A study using social robots for interviewing older adults has concluded that caregivers are generally more critical about trust in new technology than their loved ones themselves [36].

Based on these findings and Source Credibility Theory, the second hypothesis is:

**Hypothesis 2**: Participants in both the patient role and in the caregiver role score a positive usability and trust in the avatar, but in the patient role they score a higher usability and trust than in the caregiver role.

The credibility of the responses to the user questions by the avatar is also affected by the graphical design of the user interface (UI), and in particular by the avatar's visual appearance and the task domain [37]. The design of the virtual environment in which the avatar is situated, such as the chosen background and the lighting, affects the assessment of that avatar [38]. However, there are no generic design requirements for an avatar and its environment [37]. An avatar design suitable for one domain may not be suitable for another [37]. Therefore, in this study, the user interface design incorporates specially designed elements that present the avatar as an employee of a Dutch hospital located within the hospital itself. These elements aim at enhancing the user trust in the avatar. The third hypothesis is therefore as follows:

**Hypothesis 3:** User interface design elements that present the avatar as an employee of a Dutch hospital located within the hospital have a positive effect on the trust in the avatar.

## 3      Method

### 3.1      Introduction

This method section outlines the medical contexts for the two selected experimental scenarios: the first one involving participants imagining themselves as hip surgery patients, and the second one where participants imagining themselves as considering a loved one with delirium symptoms (see section 3.2). Section 3.3 covers the avatar design utilizing a large language model and retrieval augmented generation. Sections 3.4 to 3.8 detail the study design, participant selection, procedures, measurement instruments, and statistical analysis methods.

### 3.2      Total Hip Arthroplasty and Postoperative Delirium

The medical context chosen for this study is hip surgery, specifically Total Hip Arthroplasty (THA), which is a procedure with a high incidence among older adults and generally does not carry implicit negative associations unlike certain cognitive conditions. THA is one of the most frequently performed surgical interventions [44]. THA involves replacing parts of the hip joint with artificial implants. The most common cause is osteoarthritis causing hip pain [45]. The patient population receiving THA, based on OECD averages, amounts to 182 per 100,000 population [46]. The procedure of THA rehabilitation is slightly different per country. An important factor for the difference is the physical distance between the patient and the most nearby physiotherapist, especially in remote areas. The typical duration of hospital stay time is 2.6 days for elective hips, then discharge, in most cases directly to home (90%) and in some cases to a rehabilitation clinic and then home. After discharge, patients return to the hospital for control visit after 8 weeks, and after one year. Before and after surgery, several healthcare professionals (first aid doctor, nursery specialist, surgeon, physiotherapist, ...) support the patient in this "patient journey". However, they are only available for the patients a few minutes per day, whereas concerns may arise on any moment. Also, during rehabilitation, physiotherapists are only available twice a week and the remainder of the week patients have to rely on themselves. An avatar facilitating patients throughout their healthcare journey has the potential to significantly enhance patient outcomes and overall experience. Consequently, for this study, an interaction was developed in which a patient posed questions to the avatar regarding a prospective Total Hip Arthroplasty. The avatar's responses were entirely generated by the system, and a video of the interaction was produced (anonymized video 1: https://youtu.be/dN2B63gJppY).

Postoperative delirium is a frequent complication associated with THA [47]. Delirium is a severe neuropsychiatric syndrome characterized by the sudden onset of confusion, hallucinations, agitation, or other cognitive impairment [48]. The incidence of delirium in hospital patients is between 3 and 42%, and 40% in the emergency department [49]. The cause of delirium is often multifactorially determined and there are several riskfactors for its development [50, 51]. A patient's memories of the period of delirium are often not pleasant [52]. Memories of a delirious period range from detailed descriptions of perceived hallucinations to snippets of conversations with social workers and caregivers to no memories at all [53]. The patient's caregivers also experience the delirium period as particularly stressful [54]. Therefore, an avatar that enables the relatives of a patient to recognize the symptoms of delirium and provide support to the patient can be highly beneficial. Consequently, for this study, also an interaction was developed in which a relative posed questions to the avatar regarding delirium. The avatar's responses were again entirely generated by the system, and a video of the interaction was produced (anonymized video 2: https://youtu.be/mWVenC1bdy0).

### 3.3    Avatar System Design with Retrieval Augmented Generation

Figure 2 shows the avatar. The avatar has been built upon the concept described in [ref deleted for anonymization]. A central dialogue manager has been constructed in Python, which handles the speech-to-text, a query to a LLM (OpenAI API model gpt-4o-2024-08-06), and the text-to-speech functions (**Fig. 4**). Furthermore, it handles the interaction with the Unity game engine (Unity technologies, San Francisco, CA, USA) which is used as framework for the graphical user interface. This engine handles amongst others the avatar animation switching between talking and listening modes, and the lip synchronization.

The LLM-query includes Retrieval Augmented Generation (RAG), which is a technique that enhances large language models (LLMs) by incorporating external knowledge sources to improve accuracy and reduce hallucinations [55, 56]. This approach is especially useful for tasks requiring specialized knowledge beyond the LLMs training data and for prevention of hallucination [57]. An example of such an external knowledge source is the combined patient brochure set that hospitals use for informing their patients. For efficient knowledge storage and retrieval, in RAG systems large documents or document sets are often divided into chunks of typically 200-500 words [58]. Chunking methods have a significant effect on retrieval quality, which in turn influences both the response and overall user experience [44]. A pre-study conducted in our lab showed large variation in response quality for various chunking methods. Evaluating the effects of chunking were not aims of this study. Therefore, in this study, a straightforward form of Retrieval-Augmented Generation was implemented by adding one document to the large language model query. This document represents the single retrieved chunk that is explicitly included in the prompt (**Fig. 3**). The LLM is instructed to use the document as a source for answering the question and not to rely solely on its internal training.

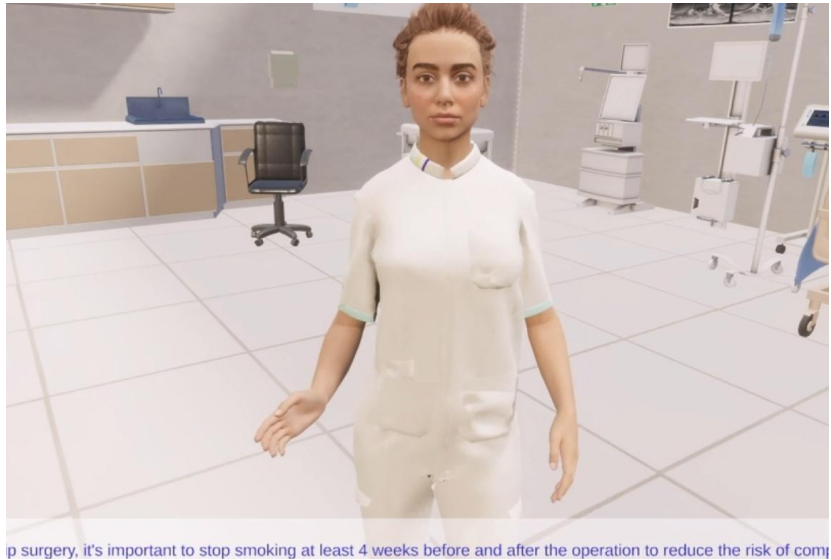

p surgery, it's important to stop smoking at least 4 weeks before and after the operation to reduce the risk of comp

**Fig. 2.** Avatar giving answer to the patient

```
try:
    if response_doc_contains_answer:
        print("Step 1 started")
        prompt = {
            f"The following document is provided: \n\n {document_text} \n\n"
            f"Please try to answer the following question in the {lang_name} language using the "
            f"information in this document {transcript}"
            f"You are not giving a list of answers, and only select the first and best answer. Use only "
            f"three sentences in your answer."
        }
        chat_completion = client.chat.completions.create(
```

**Fig. 3.** LLM Prompt

For this study, two publicly available brochures from the Canisius Wilhelmina hospital on THA and delirium in Dutch and English have been used as external knowledge source. They were converted from pdf to text file and preprocessed to remove unwanted tokens. The brochure on THA contained 6670 tokens and the brochure on delirium contained 1654 tokens. The dialogue manager (**Fig. 4**) processed the user question in two steps. First, the dialogue manager checked if the question could be answered by the information in the brochure by matching the tokens (words) in the question with the tokens in the document. If not, the general knowledge of the LLM would be used (this did not happen in the videos used in this study). If yes, the document text from the brochure was inserted into the prompt (**Fig. 3**) for the LLM with the request to answer the question using the information in the brochure.

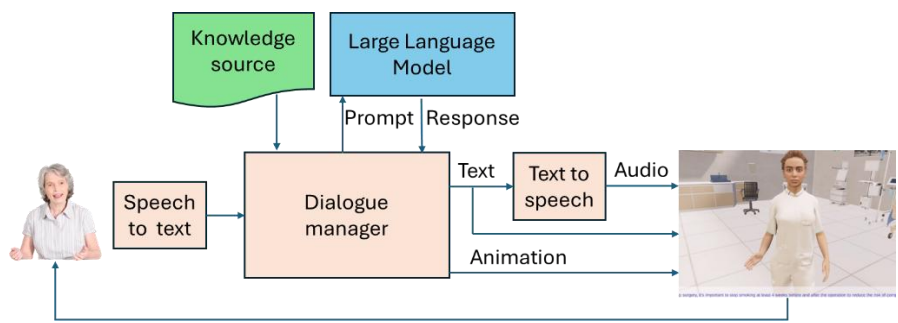

**Fig. 4.** Avatar system flow diagram

The user interface elements to be tested in hypothesis 3 include the humanoid character design created with Character Creator, whereas iClone was employed to produce the animations (both by Reallusion, San Jose, California, USA). The agent's clothing was designed to represent the uniforms worn at the Canisius Wilhelmina hospital and was created in Marvelous Designer (Seoul, South Korea). The hospital environment was based on components developed by Studio Lab (Tonbridge, United Kingdom), and modified where necessary. In the background a wall with a sign with the hospital logo could be seen. The text of both the user questions and avatar responses was shown at the bottom of the screen in separate colors in a partially transparent text box (see **Fig. 2**).

### 3.4     Study design

This study has primarily a 3x2x3 design with as independent variables the participant country of residence (Netherlands, UK or USA), gender (male or female) and education level (low, medium or high), and as primary dependent variables usability and trust. Usability and trust are 2 of the 24 constructs of the Artificial Social Agent Questionnaire (ASAQ). Usability and trust are measured by all items belonging to these constructs [59].

Secondary dependent variables were the other 22 constructs of the ASAQ questionnaire (short version), which are not measured for hypothesis testing but for exploratory reasons. This study was approved by the Ethical Committee of Radboud University under reference number ECSW-LT-2025-4-2-97936.

### 3.5     Participants

Before the experiment, a power analysis has been conducted using G*Power version 3.1.9.6 to determine the sample size [60]. The Artificial Social Agent Questionnaire (ASAQ) reports generally a small effect size (Cohen's $d$) between constructs when evaluating agents [59]. Using this small effect size of d=0.2, a power of 80% and an alpha of 0.05, for a t-test a sample size of N=156 was required. The sample in this study is N=165 and meets this requirement.

   Within the scope of this feasibility study it was not possible to invite 156 participants to the lab for a live interaction with the avatar. Therefore videos of a person interacting with the avatar were created and used in an online survey. Furthermore, these videos were intentionally kept short (1.2 and 1.3 minutes of duration, respectively) to prevent early dropout and improve experimental control.

   Participants for the survey were recruited via Prolific [61]. Inclusion criteria were being a native Dutch or English speaker, being resident in the Netherlands, the United Kingdom or the United States, and above 40 years old. The age criterion was based on the assumption that above the age of 40 it is more likely that one is concerned about a THA for a relative or oneself. The included sample has a mean age of 51.6 ($SD$=8.9) and consisted of 80 males and 85 females. There were no participants who identified themselves with another gender.

### 3.6     Procedure

Participants could volunteer from their digital device by clicking a link on the Prolific platform which directed them to a survey designed in LimeSurvey [62]. First they were required to provide informed consent. Then they were asked to provide their demographic details, i.e. age, gender and education level. For each country, the education levels were categorized into country specific groups that can be classified as low, medium and high.

The fact that the avatar used in the background a LLM enhanced with the RAG-technique including a hospital provided document for answering the user questions, cannot be seen from the user interface. Therefore, participants were presented with an introductory text stating that the avatar would respond to questions using information provided by the hospital.

Next a short video of 1.2 minutes was shown, where the participant could see and hear the avatar responding to spoken questions from a typical patient concerning the Total Hip Arthroplasty. After the video, participants were asked to score statements about the avatar from a patient perspective on a 7-point Likert scale. Then, a second short video of 1.3 minutes was shown, where the participant could see and hear the avatar responding to typical questions from a relative concerning delirium. Next, the participants were asked to score additional statements about the avatar from the perspective of a relative, also on a 7-point Likert scale. The videos were not presented in random order, as each situation is distinct and not intended for direct comparison. Finally, participants were thanked for their participation and redirected to the Prolific site. Participants received a compensation related to the time needed for completing the task (on average 7.58 minutes) of €12.- per hour. The videos were available in the Dutch language for the Dutch population and in the English language for the UK and USA populations.

### 3.7    Measurement Instruments

The agent's usability was measured using the three items of the ASAQ questionnaire construct 'Agent's usability' [59]. These items are 'The agent is easy to use', 'Learning to work with the agent is easy', and 'Learning how to communicate with the agent is quick'. The agent trust was measured using the three items of the ASAQ questionnaire construct 'User's Trust'. These items are 'The agent always gives good advise', 'The agent acts truthfully', and 'I can rely on the agent'. Construct scores for usability and trust were calculated by averaging the item scores. Participants were asked to score these two constructs with three items after each video, one time from patient perspective (video 1), and the second time from the perspective of a relative (video 2). To limit the participant burden, the other 22 ASAQ constructs were only measured for exploratory reasons after the first video and using only the main item of the construct as indicated in the ASAQ short version.

The evaluation of virtual humans from a user interface perspective has followed the initial steps defined in  the Interaction Design textbook by Rogers, Sharp and Preece [63]. Participants were asked to evaluate five specific items in the user interaction that were incorporated in the UI to enhance trust. These items concern the use of information from the hospital brochures, the inclusion of the hospital logo in the background (cutoff for anonymity in **Fig. 2**), the presentation of the answer in text at the bottom of the UI screen (in addition to the spoken text), the depiction of the agent in the hospital's nurse uniform, and the portrayal of the agent against a typical hospital background.

### 3.8    Data analysis

Data from the surveys were downloaded from the LimeSurvey platform and imported into SPSS version 30.0.0 for further statistical analysis. Data from incomplete surveys were deleted, and irrelevant data were removed. Hypothesis 1 was evaluated with three-way ANOVAs per dependent variable and post-hoc Tukey HSD tests if needed. Hypothesis 2 was evaluated using descriptive statistics and a t-test. Hypothesis 3 was evaluated with descriptive statistics.

## 4      Results

### 4.1    Effects of country of residence, gender and education level

The participant sample was located in the Netherlands (n=47), the UK (n=86), and the USA (n=32). Among them, 85 identified as female and 80 as male. Regarding education level, 122 participants reported a high education level, 40 reported a medium education level, and 3 reported a low education level.

For participants imagining themselves as a patient in the Total Hip Arthroplasty scenario, a 3 (Country of Residence: NL, UK, US) x 2 (Gender: male, female) x 3 (Education Level: low, medium, high) ANOVA was conducted on test scores for usability. There were no significant main effects of Country of Residence, $F(2, 162) = 0.65$, $p = .52$, $\eta^2 = .01$, Gender, $F(1,163) = 0.05$, $p = .83$, $\eta^2 < .001$, or Education Level, $F(2, 162) = 1.04$, $p = .36$, $\eta^2 = .01$. In addition, none of the two-way interactions nor the three-way interaction were significant, all $Fs < 1.56$, $ps > .21$.

For the same participants imagining themselves as a patient, another three way-ANOVA was conducted on test scores for trust. There were no significant main effects of Country of Residence, $F(2, 162) = 0.12$, $p = .89$, $\eta^2 = .002$, Gender, $F(1,163) = 0.003$, $p = .96$, $\eta^2 < .001$, or Education Level, $F(2, 162) = 0.08$, $p = .92$, $\eta^2 = .001$. Also for trust, none of the two-way interactions nor the three-way interaction were significant, all $Fs < 0.38$, $ps > .80$.

Next, for participants imagining themselves as a caregiver in the post-operation delirium scenario, also a 3 (Country of Residence: NL, UK, US) x 2 (Gender: male, female) x 3 (Education Level: low, medium, high) ANOVA was conducted on test scores for usability. There were no significant main effects of Country of Residence, $F(2, 162) = 0.96$, $p = .39$, $\eta^2 = .01$, Gender, $F(1,163) = 0.004$, $p = .95$, $\eta^2 < .001$, or Education Level, $F(2, 162) = 0.82$, $p = .44$, $\eta^2 = .01$. Additionally, none of the two-way interactions nor the three-way interaction were significant, all $Fs < 1.52$, $ps > .22$.

Finally, for the same participants imagining themselves as a caregiver, the final three way-ANOVA was conducted on test scores for trust. There were no significant main effects of Country of Residence, $F(2, 162) = 0.48$, $p = .62$, $\eta^2 = .006$, Gender, $F(1,163) = 0.473$, $p = .49$, $\eta^2 = .003$, or Education Level, $F(2, 162) = 0.15$, $p = .86$, $\eta^2 = .002$. Furthermore, none of the two-way interactions nor the three-way interaction were significant, all $Fs < 1.49$, $ps > .23$.

These results indicate that hypothesis 1 can be accepted.

### 4.2    Usability and trust across complete groups

Participants (N=165) imagining themselves as a patient in the Total Hip Arthroplasty scenario score the usability (*M*=4.606, *SD*=0.967) and trust (*M*=3.923, *SD*=1.104) in the avatar both positive, i.e. > 3.

Participants (N=165) imagining themselves as a relative in the post-operative delirium scenario also score the usability (*M*=4.638, *SD*=1.079) and trust (*M*=4.309, *SD*=1.052) both positively (>3).

When comparing the usability between both scenarios, Levene's test for equality of variances showed no significant difference in variances. A within-subject t-test showed no difference between the usability scores *t*(164) = -.682, *p* = .248.

When comparing the trust between both scenarios, Levene's test for equality of variances also showed no significant difference in variances. However, participants scored the trust in the avatar lower when imagining themselves as a patient than as a relative *t*(164) = -7.787, *p* < .001, *d* = .636, 95% *CI* [-0.772, -0.439].

Hypothesis 2 can be partly accepted.

### 4.3    Effects of User Interface Elements across complete groups

The descriptive statistics of the contribution of the user interface elements are provided in **Table 1**. Most elements contribute to the factor trust, with a score > 3, and with the presentation of the text as the highest contributor. The results in this table show that hypothesis 3 can be accepted.

| Factor | Mean | SD |
|---|---|---|
| Information from the hospital brochures | 3.887 | 1.377 |
| Inclusion of the hospital logo in the background | 3.521 | 1.496 |
| Presentation of the answer in text at the bottom of the UI screen | 4.418 | 1.482 |
| Depiction of the agent in the hospital's nurse uniform | 3.430 | 1.424 |
| Portrayal against a typical hospital background | 4.012 | 1.423 |

**Table 1.** User Interface screen elements contibuting to trust

### 4.4    Other human interaction experiences

The following **Table 2** reports the exploratory data on the other 22 ASAQ human interaction experience constructs for all participants imagining themselves as a patient in the hip surgery scenario (N=165). The usability and trust constructs in accordance with the ASAQ short questionnaire (thus measured with 1 item instead of 3) are also included for sake of completeness. Items were reversed when needed and are sorted on mean value in **Table 2**.

| Code | Construct | M | SD |
|---|---|---|---|
| AU | Usability | 4.539 | 1.062 |
| PF | Performance | 4.249 | 1.212 |

| AI | Intentionality | 4.012 | 1.728 |
|-----|-----------------|-------|-------|
| AC | Coherence in behaviour | 3.988 | 1.506 |
| AAS | Appearance suitability | 3.939 | 1.464 |
| AA | Attentiveness | 3.691 | 1.417 |
| UT | User Trust | 3.649 | 1.422 |
| UE | User's engagement | 3.485 | 1.614 |
| AT | Attitude | 3.394 | 1.549 |
| APP | Personality presence | 2.879 | 1.561 |
| AE | Enjoyability | 2.849 | 1.564 |
| AL | Likeability | 2.818 | 1.668 |
| HLA | Human-like appearance | 2.703 | 1.593 |
| UAA | User acceptance | 2.649 | 1.707 |
| UEP | User's emotion presence | 2.600 | 1.637 |
| IIS | Interaction impact on self-image | 2.527 | 1.599 |
| UAI | User-agent interplay | 2.461 | 1.702 |
| AS | Sociability | 2.376 | 1.624 |
| HLB | Natural behaviour | 2.230 | 1.591 |
| UAL | User-agent alliance | 2.170 | 1.521 |
| NA | Natural appearance | 2.121 | 1.501 |
| SP | Social presence | 1.861 | 1.630 |
| AEI | Emotional intelligence presence | 1.527 | 1.360 |

**Table 2.** ASAQ constructs

A mean value > 3 indicates a positive contribution to human interaction experience. From the top five construct scores, it appears that the avatar is easy to use, does its task well and knows what it is doing. Its behaviour is coherent and suitable for the task. From the lowest five construct scores, it appears however also that the avatar does not seem natural from its behaviour and outer appearance, is not considered as a social entity, and is considered emotionless. The avatar does not form a natural alliance with the user.

## 5    Discussion and conclusion

The usability of the avatar for participants imagining themselves in the patient role was evaluated positively. The video showed a scenario in which a patient asks questions to the avatar and participants seem to consider this scenario useful and trustworthy. They may be able to imagine that they themselves could be such a patient talking to the avatar. This is consistent with most findings in a systematic review by Milne-Ives et al. (2020) on thirty-one studies that evaluate several forms of conversational agents [64]. It should be noted that in their systematic review, only six out of the thirty-one studies included some form of voice interaction, and none concerned the combination of avatar, LLM and RAG [64].

The usability of the IVA for participants imagining themselves in the caregiver role for the case of delirium was also evaluated positively. This is also consistent with the studies referenced in the work of Milne-Ives et al. (2020) [65].

Regarding trust, the Source Credibility Theory explained that competence is a major factor in building trust [66]. This trust was also evaluated positively in both scenarios. The system design elements, such as the use of brochure information and the convenient display of spoken text (**Table 1**), may have contributed to competence and trust levels, and this is in line with several factors identified by Rheu et al. [67]. However, participants scored trust in the avatar lower when imagining themselves as patients than when imagining themselves as relatives, and this was contrary to expectations. A possible explanation might be the difference between the scenarios. Participants asked to imagine being a patient were told to consider themselves scheduled for Total Hip Arthroplasty, which is not difficult to envision undergoing personally. Participants may have found it more challenging to imagine being a relative of a patient with post-surgical delirium, as this symptom may be harder to envision. Further research is needed to explain this outcome.

Based on the Social Identity Theory, no differences among experiment subgroups divided by country of residence, gender, or education level were predicted. This prediction was confirmed by the results. This could be explained by the fact that the Netherlands, the UK and the USA have high levels of health care and participants seem to judge the provided health care similarly. Also, they could imagine being supported by this type of technology. Similarly, no differences were found between males and females, and these results suggest that both have a similar opinion about the usefulness of avatars. Regarding education level, it should be noted that the groups where strongly skewed, with only 3 having a low education and 122 having a high education. A follow-up study should aim for more participants with a low education level.

The results suggest that UI elements may increase trust among people over 40. The presentation of the answer in text at the bottom of the UI screen contributed most to ths factor trust. This could be explained by the audio possibly not being always clearly audible, and the display of the text helps to understand what is said by the avatar. The depiction of the agent in the hospital's uniform seems to have contributed the least to the trust, although it was still positive. It should be noted that the creation of the uniformed character turn out to be quite complex. Many trials were required to create the character in Character Creator, import it in Marvelous Designer to design the uniform, export it back to Character Creator, then export it to fbx-format and import it in Unity, without losing parts of the clothing or getting an unrealistically moving uniform during animations. All factors require more research to establish which elements have the most impact.

This study has several limitations. Firstly, participants were unable to interact with the system directly. Organizing live interactions with the avatar for all 156 participants required by the power analysis was not feasible in this study. Although video-based evaluation offers more experimental control, it significantly limits the assessment of the RAG based LLM avatar's interactive capabilities, which is important for evaluating its practical effectiveness and reliability. Experiments with an actual interaction are important for more reliable results. A live interaction study is also a better opportunity to compare a RAG supported LLM system with a vanilla LLM. Secondly, the videos were short; longer videos could showcase more capabilities but risk early participant dropout. Furthermore their presentation order was not randomized, which could introduce potential order effects that confound the results. At the same time, both situations were quite different and it was not the intention to compare both scenarios in the sense of A/B test. Thirdly, the sample was skewed towards higher educated individuals. Some studies indicate that conversational agents can improve health information access for those with low health literacy [3]. A more equitable distribution of education levels among participants may yield improved outcomes for the group with low health literacy. Finally, under the hood of the interface screen, technical enhancements can be implemented concerning RAG techniques as well as concerning speech-to-text and text-to-speech functionalities.

The study's consistent results across participant countries suggest generalizability to nations with similar healthcare systems and living standards. Gender consistency indicates potential generalizability in terms of gender. Further research is needed to confirm this.

Moreover, future studies should involve extended and live interactions with the system, improved graphical interface quality, better voice control, and enhanced character facial and bodily animations.

This study did not explicitly address privacy issues. The avatar in the videos answered questions that did not contain personal information and used public hospital information material. However, it is likely that patients or relatives would share private information on purpose or inadvertently at some stage in the interaction. Future research should study solutions to these problems, such as using local LLMs.

Although this study indicates that LLM queries are useful, they also have a carbon footprint. Tomlinson et al. (2024) report that combined impact of training and operation for ChatGPT amounts to approximately on average 2.2 g $CO_2$ per query [68]. Future research should focus on identifying optimal strategies to minimize or offset this footprint, while ensuring that patient and caregiver needs for accessible and reliable hospital information are met.

The preliminary findings of this study cautiously suggest that an avatar on the Computer-On-Wheels platform may have potential for providing patients with needed medical information during hospital stays, thanks to the Retrieval-Augmented Generation enhanced Large Language Model. However, further research is necessary to confirm and expand upon these results.